# Relationship between the Uncompensated Price Elasticity and the Income Elasticity of Demand under Conditions of Additive Preferences

Lorenzo Sabatelli, PhD

GLOBMOD Health, Market Analysis Unit, Barcelona, Spain

lorenzo.sabatelli@globmod.com

Keywords: consumer choice, demand, income elasticity, price elasticity, preference independence, forecasting, mathematical model, microeconomics, pricing, financial instruments of policy

Abstract

Income and price elasticity of demand quantify the responsiveness of markets to changes in income and in prices, respectively. Under the assumptions of utility maximization and preference independence (additive preferences), mathematical relationships between income elasticity values and the uncompensated own and cross price elasticity of demand are here derived using the differential approach to demand analysis. Key parameters are: the elasticity of the marginal utility of income, and the average budget share. The proposed method can be used to forecast the direct and indirect impact of price changes and of financial instruments of policy using available estimates of the income elasticity of demand.

Introduction

A change in the price of a market good determines a change in the purchasing power of consumers (income effect), and a change in the relative price of goods (substitution effect). The aggregate consumer responsiveness to changes in price and in income is measured using the (own and cross) price elasticity and the income elasticity of demand, respectively.

Knowing the uncompensated own and cross price-elasticity of demand is essential to anticipate the impact of price changes and of financial instruments of policy, such as subsidies, cost-sharing schemes, and taxation, nonetheless forecasting it requires data that is not always readily available. Contingency studies, e.g. willingness-to-pay studies, are often used to elicit the potential response of consumers, nevertheless they are not always financially and logistically feasible, or consistent [1][2]. Unlike the price elasticity, the income elasticity of demand can often be estimated from routinely collected data (e.g. from household surveys), and is therefore more readily available. Nonetheless, it does not contain in and of itself enough information to infer the consequences of changing prices.

The mathematical relationship between demand and price can be modeled using the neoclassical consumer theory, assuming a representative economic agent with preferences over consumption goods, captured by a utility function [3]. The Rotterdam model, first proposed by Barten (1964) [4] and Theil (1965) [5], builds on this approach, allowing for the estimation of substitutes and complements, and separability of preferences. The Rotterdam model produces constant marginal shares, a problem that can be avoided using a demand function called the almost ideal demand system (AIDS) model [6], which was subsequently extended by Theil, Chung, and Seale [7][8]. They added a non-linear substitution term to the basic linear function, which allows for separability and has fewer parameters to be estimated than in the AIDS model, creating the Florida model [7][8]. If separability holds, total expenditure can be partitioned into groups (or bundles) of goods, making it possible to analyze the preferences for one group independently of other groups. In that case, the mathematical relationship between price and demand becomes amenable to analytical calculations.

In the present study, mathematical relationships that allow the estimation of the uncompensated own price elasticity and of the cross price elasticity of demand for independent bundles of goods are obtained following the differential approach used to derive the Florida model. The proposed equations require three inputs: the income-elasticity of demand, the mean budget share allocated to the bundle of goods of interest, and the elasticity of the marginal utility of income.

## Methods

Relationship between Income Elasticity and Price Elasticity of Demand

The definitions used throughout this paper are reported in Table 1. The following assumptions are made:

I. The utility function is strictly concave twice continuously differentiable (i.e. the Hessian matrix is continuous and negative definite);

II. The consumers have a limited budget and they allocate it in a way that maximizes individual utility;

III. Preference independence: the utility generated by the consumption of a bundle of goods does not depend on the consumption of goods from other bundles. In other words, the utility is the sum of the utilities associated with the consumption of each individual bundle of goods;

IV. The elasticity of the marginal utility with respect to income is constant.

Theorem: Given the assumptions I-II-III-IV, the mean value of the budget share ($\omega$) spent on each bundle, and the elasticity of the marginal utility of income ($\rho$), we show that for a given bundle of goods (*i*), a quantitative (parabolic) functional relationship exists between the income elasticity of demand ($\varepsilon$), and the uncompensated own price elasticity of demand ($\eta$):

$$\eta_i = -\frac{1}{\rho}\omega_i\varepsilon_i^2 + \left(\frac{1}{\rho} - \omega_i\right)\varepsilon_i \qquad (1)$$

and that the cross-price elasticity ($\psi$) of demand for a bundle of goods (*i*) with respect to the price of a bundle (*j*) is:

$$\psi_{ij} = -\frac{1}{\rho}\omega_j\varepsilon_i\varepsilon_j - \omega_j\varepsilon_i \qquad (2)$$

The parameter $\rho$ is estimated analyzing surveys of subjective happiness [9], and its value appears to be quite stable across different geographic areas and populations groups, with an average value equals to -1.26 and a standard deviation equals to 0.1. $\omega$ is estimated from household surveys, and from standard market-research data. To take into account the effect of parametric uncertainty on model estimates, credible intervals for the estimates of the price elasticity of demand are calculated via Monte-Carlo simulation, drawing random model parameter values from normal (*N*) and uniform (*U*) distributions::

$$\rho \sim N(\mu_\rho, \sigma_\rho)$$
$$\omega_i \sim U(\omega_{min}, \omega_{max})$$
(3)

The Proof

The derivation of Eq. (1) and Eq. (2) uses Lagrange multipliers and differential equations, and is based on the fact that any change in the price of a good determines a change in the purchasing power of the consumer (income effect), and a change in the relative price of goods (substitution effect). The substitution effect depends on two elements: a) the relative importance of different goods to the consumer; and (b) the deflationary impact that a change in the price of a single good has on all market goods. The proof follows three steps:

1. A demand equation for a bundle of goods (*i*) is derived using the Theil's [7] and Barten's [4] approach;

2. Analytical expressions for income elasticity, uncompensated own price elasticity, and cross price elasticity of demand for bundle (*i*) are derived from the demand equation, under the assumption of preference independence;

3. The analytical expressions obtained in step-2 are then combined to derive Eq. (1) and Eq. (2).

Step 1

Let's first define the vector of the quantities of goods that the consumer purchases from each bundle:

$$\bar{q} = (q_1, q_2..., q_n)$$

and the vector containing the average prices paid for the goods in each bundle :

$$\bar{p} = (p_1, p_2..., p_n).$$

The following vector notation is used:

$$\frac{d}{d\bar{q}} = \left[ \frac{\partial}{\partial q_i}, ............, \frac{\partial}{\partial q_n} \right]$$

$$\bar{q} \bullet \bar{p} = \sum_{i=1}^{n} q_i p_i$$

Assumption (I) implies that first and second order derivatives of the utility function $u$ exist, and that the Hessian matrix of $u$ is symmetric negative.

Assumption II implies that:

- Given the budget constraint

$$E = \bar{p} \bullet \bar{q} \quad (4)$$

The following Lagrangian function can be defined:

$$F = u(\bar{q}) - \mu(\bar{p} \bullet \bar{q} - E) \quad (5)$$

- $u$ can be maximized, subject to the budget constraint (4), by using the Lagrangian multiplier method, which consists in searching the values of $\bar{q}$ such that the gradient of $F$ is null and the Hessian of $F$ is negative defined. Differentiating $u$ with respect to $\bar{q}$ :

$$\frac{du}{d\bar{q}} = \mu \bar{p} \quad (6)$$

Combining (6) with the budget constraint (4), and following the Barten's approach [4] the result can be written in the following partitioned matrix form:

$$\begin{bmatrix} U & \bar{p} \\ \bar{p}^T & 0 \end{bmatrix} \begin{bmatrix} dq/dE & dq/dp^T \\ -d\mu/dE & -d\mu/dp^T \end{bmatrix} = \begin{bmatrix} 0 & \mu I \\ 1 & -q^T \end{bmatrix} \quad (7)$$

Where $I$ is the identity matrix.

Solving the matrix demand equation, Eq. (7), and following Theil's derivation [7], the general form of the differential demand system is obtained:

$$\omega_i d\log q_i = \theta_i d(\log Q) + \frac{1}{\rho} \sum_{j=1}^{n} \theta_{ij} d\left[\log \frac{p_j}{p^*}\right] \quad (8)$$

where $\theta$ is the marginal share, and the following equation holds for the (Frisch) price deflator:

$$d(\log p^*) = \sum_{i=1}^{n} \theta_i \frac{dp_i}{p_i} \quad (9)$$

The first term on the right of Eq. (8) is the real income term of demand, which results from the change in money income and the income effect of the price change. The second term on the right of Eq. (8) is the substitution term. Under preference independence (assumption III), the term containing the Frisch deflated price of good $i$ is the only non-zero term in the substitution term. Therefore Eq. (8) becomes:

$$\omega_i d\log q_i = \theta_i d(\log Q) + \frac{1}{\rho} \theta_i d\left[\log \frac{p_i}{p^*}\right] \quad (10)$$

Step 2

Differentiating the budget constraint (4) yields:

$$d(\log E) = d(\log Q) + d(\log P) \quad (11)$$

where:

$$d(\log Q) = \sum_{i=1}^{n} \omega_i \frac{dq_i}{q_i} \qquad (12)$$

$$d(\log P) = \sum_{i=1}^{n} \omega_i \frac{dp_i}{p_i} \qquad (13)$$

If prices are kept constant, the following equation holds:

$$d(\log E) = d(\log Q) \qquad (14)$$

Since $\rho$ does not depend on income (assumption IV), the income elasticity of demand $\varepsilon$ is:

$$\varepsilon_i = \frac{d(\log q_i)}{d(\log E)} = \frac{\theta_i}{\omega_i} \qquad (15)$$

which is obtained substituting Eq. (14) in Eq. (10), and then differentiating with respect to $d(\log E)$.

The uncompensated own price elasticity of demand $\eta$ can be derived substituting Eq (9) in Eq. (10), and using the equation:

$$d(\log Q) = d(\log E) - d(\log P) = d(\log E) - \sum_{i=1}^{n} \omega_i \frac{dp_i}{p_i}$$

Since $\rho$ does not depend on prices (assumption IV), and given Eq. (9), $\psi$ is given by:

$$\psi_{ij} = \frac{d(\log q_i)}{d(\log p_j)} = -\frac{1}{\rho} \frac{\theta_i}{\omega_i} \theta_j - \frac{\omega_j}{\omega_i} \theta_i + \frac{1}{\rho} \frac{\theta_i}{\omega_i} \delta_{ij} \qquad (16)$$

where:

$$\delta_{ij} = \begin{cases} 1 & \text{if } i = j \\ 0 & \text{if } i \neq j \end{cases}$$

Step 3

Obtaining the values of $\theta$ from Eq. (15) and substituting them into Eq. (16) yields:

$$\psi_{ij} = -\frac{1}{\rho}\omega_j\varepsilon_i\varepsilon_j - \omega_j\varepsilon_i + \frac{\delta_{ij}}{\rho}\varepsilon_i \quad (17)$$

Eq. (1) is equivalent to Eq. (17) for $i=j$:

$$\eta_i = -\frac{1}{\rho}\omega_i\varepsilon_i^2 + \left(\frac{1}{\rho} - \omega_i\right)\varepsilon_i$$

and Eq. (2) is equivalent to Eq. (17) for $i \neq j$:

$$\psi_{ij} = -\frac{1}{\rho}\omega_j\varepsilon_i\varepsilon_j - \omega_j\varepsilon_i$$

As an example, the derivation of Eq. (1) for a hypothetical market trading only in two independent bundles of goods is outlined in the S1.text.

## Results and Discussion

Analytical calculations show that the uncompensated own price elasticity $\eta$ and cross price elasticity of demand $\psi$ of independent bundles of goods can be expressed as functions (Eq. (1) and Eq. (2)) of the income elasticity of demand, the average budget share $\omega$, and the elasticity of marginal utility of income $\rho$:

$$\eta_A = -\frac{1}{\rho}\omega_A \varepsilon_A^2 + \left(\frac{1}{\rho} - \omega_A\right)\varepsilon_A$$

$$\psi_{AB} = -\frac{1}{\rho}\omega_B \varepsilon_A \varepsilon_B - \omega_B \varepsilon_A$$

For instance, for a bundle of goods A, if $\omega$ is in the range 0.01%-10%, $\rho$ is drawn from a normal distribution with mean equal to -1.26 and standard error equal to 0.1, and $\varepsilon$ is equal to 1, then Eq. (1) predicts a price elasticity equal to -0.8 (with the 95% credible interval: -0.96, -0.64). The statistical error associated with the price elasticity estimates increases with the value of the income elasticity (see Fig 1). The sensitivity to model parameter values (and especially to the value of the elasticity of the marginal utility of income) increases with the income elasticity of demand, and therefore so does the width of credible intervals associated with predictions. A univariate sensitivity analysis (whose results are not displayed), indicates that the price elasticity would change by less than 7% (relative change), as the average budget share fluctuates between 0.01% and 10%, suggesting that for bundles of goods that account for less than 10% of the total expenditure an approximate estimate of the average budget share may be sufficient to produce relatively accurate estimates of the own price elasticity.

According to Eq. (2), the cross price elasticity of demand for a bundle of goods A with respect to a bundle B grows linearly with the income elasticity of demand for A (Fig 2). The slope depends on the income elasticity of demand for B, on the budget share of B, and on the elasticity of the marginal utility of income. If the income elasticity of B is smaller than the absolute value of the elasticity of the marginal utility of income, an increase in the price of B will determine a reduction in the demand for A. If the income elasticity of B is larger than the absolute value of the elasticity of the marginal utility of income, an increase in the price of B will determine an increase in the demand for A. This is the effect of a reallocation of the budget performed by the consumer to offset the consequences of a change in the price of B on real income and on the relative prices of market goods. The statistical uncertainty associated with cross price elasticity estimates increases with the income elasticity of B (see Figs 2 and 3), and (unlike the case of the own price elasticity) is rather sensitive to the actual value of the budget share.

Eq. (1) can be used to forecast the impact of a change in the own price of bundle A on the demand for A, while Eq. (2) can be used to assess how the demand for a bundle A changes if the price of an (independent) bundle B changes. As such, they provide a theoretical basis for estimating the potential impact of financial instruments of policy, e.g. subsidies and cost-sharing schemes, using household survey data [10] [11], and for scenario-analysis in early stage product pricing, if the relevant target is a bundle of goods that can be treated as preference independent. The face validity of this approach was tested using publicly available data. Model predictions from Eq. (1) were compared with published estimates of income and price elasticity for different bundles of goods and services that had been calculated fitting the Florida model to real world data collected in national surveys of consumption [8]. To ensure comparability, bundles accounting for an average budget share smaller than 10% were selected. Specifically, the bundles considered were: clothing and footwear, education, healthcare, and recreation. Average elasticity values for low-, middle-, and high-income countries are displayed in Fig 4. Country-specific data can be found in S2Fig. The results show that the data are confined within the (funnel-shaped) 95% credible interval region of model predictions, close to the median value curve.

Nevertheless, there are several assumptions and caveats that need careful consideration. The accuracy of estimates based on the model here presented may be limited by the assumption of instantaneous maximization of consumer utility, which in practice requires maximization of utility to be performed in a relatively short time. The proposed approach relies on strong separability assumptions which are often not met in real markets. In addition, preferences exhibit non-satiation, i.e. goods are assumed to be available in all quantities, and a consumer may choose to purchase any quantity of a good she desires. Whilst these assumptions allow a useful simplification of the calculations involved, they might also be sources of bias in the results.

Nonetheless, similar caveats do also apply to other models [7][8][12][13]. In fact, the mathematical framework here described builds on approaches already used in those studies, but with different objectives. For instance, Barnett and Serletis [13] used the differential approach to demand analysis, and then implemented the Rotterdam parameterization to move from a model based on infinitesimal instantaneous changes to a discrete model, where prices, income, and demand change over finite time intervals (days, months, years). Subsequently they fitted the discrete model to time series of prices and income, to calculate the parameter values of the demand system. Brown and Lee (2002) [12] followed an approach often used when modeling advertising effects in the Rotterdam model, and introduced preference variables in the utility function. They then studied the effect of imposing restrictions on preference variables. Nevertheless, to-date no published study has investigated the analytical relationship between income elasticity and price elasticity of demand. When compared with the results of studies that have concurrently estimated price and income elasticity of demand [8][14][15][16], the predictions of the model here presented appear consistent with the available data (see Fig 4, S1.Fig, and S2.Fig).

In conclusion, based on theoretical considerations and on the available evidence, the estimates of price elasticity of demand obtained using the analytical relationships here proposed are comparable with those generated using other models based on the differential approach to demand analysis. If used to infer the price elasticity from available estimates of the income elasticity of demand, under conditions of additive preferences, the proposed model provides an effective shortcut to forecast the impact of price changes on consumption patterns.

Acknowledgments

The author wishes to thank Prof. Dean T. Jamison (UCSF) for useful discussions. The author acknowledges partial financial support from the Bill and Melinda Gates Foundation through the Disease Control Priorities Project (Department of Global Health of the University of Washington, Seattle).

References

1. Cameron TA. Contingent valuation. The New Palgrave Dictionary of Economics Online 2ed; 2008.

2. Knetsch JL, Sinden JA. Willingness to Pay and Compensation Demanded: Experimental Evidence of an Unexpected Disparity in Measures of Value. The Quarterly Journal of Economics 1984; 99 (3): 507-521. The MIT Press.

3. Deaton A. Consumer expenditure The New Palgrave Dictionary of Economics 2ed; 2008.

4. Barten AP. Consumer Demand Functions Under Conditions of Almost Additive Preferences. Econometrics 1964; 32: 1-38.

5. Theil H. The Information Approach to Demand Analysis. Econometrica 1965, 33: 67-87.

6. Deaton A, Muellbauer J. An Almost Ideal Demand System, The American Economic Review 1980; 70 (3): 312–326.

7. Theil H. Chung C-F, Seale, JL International evidence on consumption patterns. JAI Press, Inc. Appendix B:The Differential Approach to Consumption Theory; 1989.

8. Seale JL, Regmi A, Bernstein J. International evidence on food consumption patterns, Economic Research Service, US Department of Agriculture; 2003.

9. Layard R, Mayraz G, Nickell S. The marginal utility of income. Journal of Public Economics 2008; 92: 1846-1857.

10. Sabatelli L, Jamison, DT. A model using household income and household consumption data to estimate the cost and the effectiveness of subsidies: a modelling study using cross-sectional survey data. The Lancet 2013; 381: S128.

11. Sabatelli L. A. Methodology for Predicting the Impact of Co-payments on the Utilization of Health Technologies. Value in Health 2013; 16(7).

12. Brown MG, Lee J-Y. Journal of Agricultural and Applied Economics 2002; 34: 17-26.

13. Barnett WA. Serletis A. The differential approach to demand analysis and the Rotterdam model. Quantifying consumer preferences, Contributions to Economic Analysis. Emerald Group Publishing Limited, 61-82; 2009.

14. Manning W, Newhouse J, Duan N, Keeler E, Leibowitz A. (1987) Health insurance and the demand for medical care: evidence from a randomized experiment. The American Economic Review 1987; 77 (3): 251–277.

15. Ringel, JS, Hosek SD, Vollaard BA. Mahnovski S. The Elasticity of Demand for Health Care: A Review of the Literature and Its Application to the Military Health System; Santa Monica, CA: RAND Health; 2002.

16. Santerre RE, Vernon JA. Assessing Consumer Gains From A Drug Price Control Policy In The United States. Southern Economic Journal 2006; 73: 233-245.

Figures

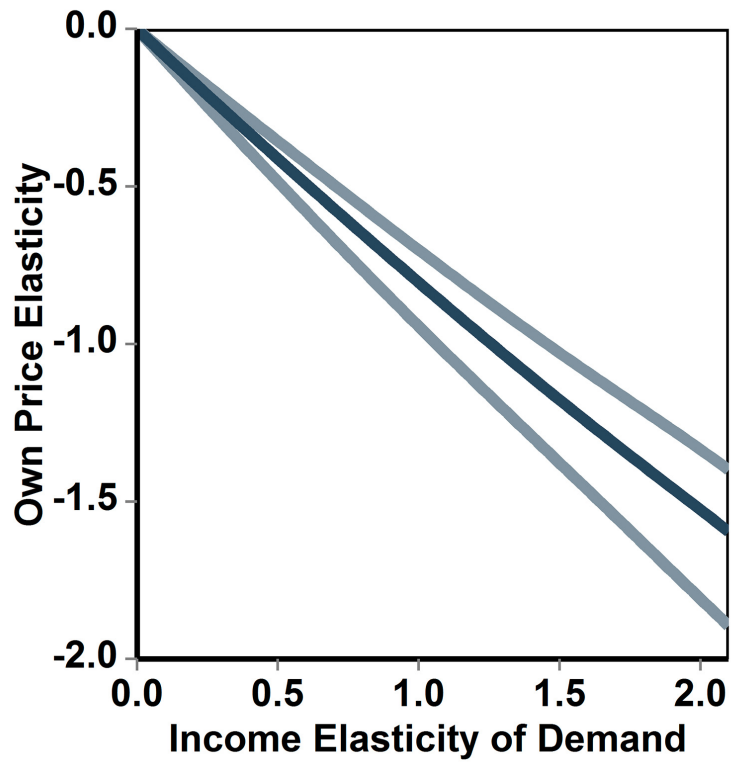

Fig 1. Relationship between income elasticity and uncompensated own price elasticity of demand. The darker line (in the middle) indicates the median of simulated values, while the lighter external lines define the 95% credible interval calculated using a Monte-Carlo simulation. The average budget share was drawn from a uniform distribution ranging from 0.0001 to 0.1, and the elasticity of the marginal utility of income was drawn from a normal distribution with mean equal to -1.26 and standard deviation equal to 0.1.

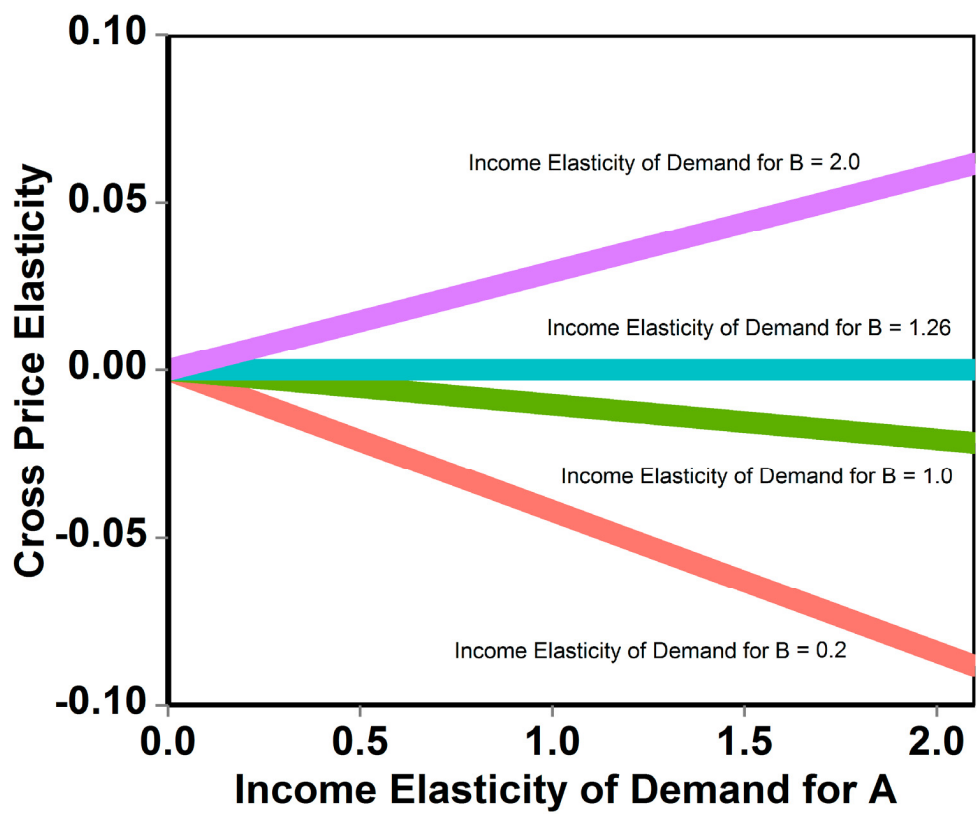

Fig 2. Relationship between income elasticity of two (preference) independent bundles of goods A and B, and the cross price elasticity of demand for a bundle of goods A with respect to B. The cross price elasticity is negative, null or positive, depending on whether the income elasticity of B is smaller of, equal to, or larger of the absolute value of the elasticity of the marginal utility of income. The average budget share is equal to 0.05 and the elasticity of the marginal utility of income is equal to -1.26.

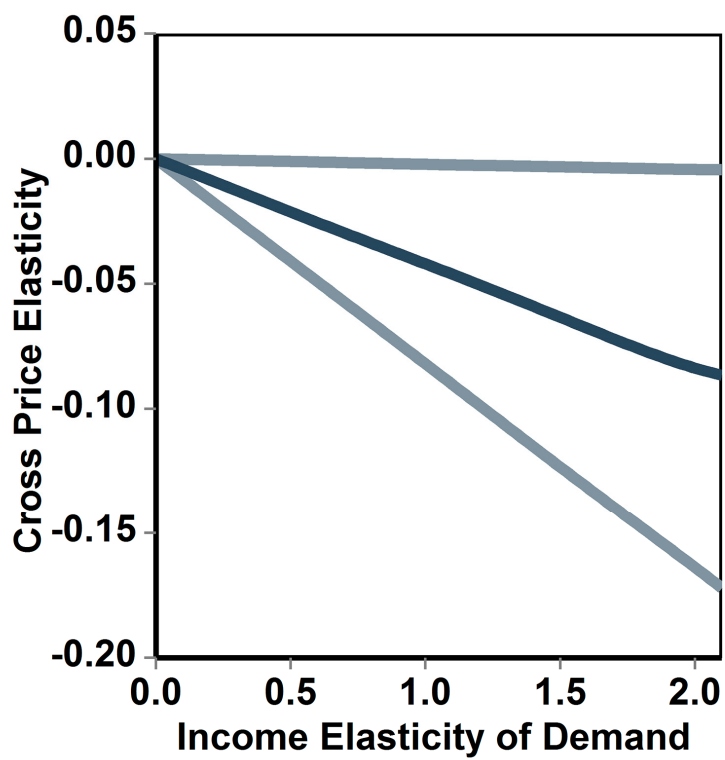

Fig 3. Quantification of the uncertainty affecting the estimates of the cross price elasticity of demand for a bundle of goods A, whith respect to the price of B. In this example, the income elasticity of the bundle of goods B is equal to 0.2, and the cross price elasticity is plotted against the income elasticity of demand for A. The darker line (in the middle) indicates the median of simulated values, while the lighter external lines define the 95% credible interval calculated using a Monte-Carlo simulation. The average budget share was drawn from a uniform distribution ranging from 0.0001 to 0.1, and the elasticity of the marginal utility of income was drawn from a normal distribution with mean equal to -1.26 and standard deviation equal to 0.1.

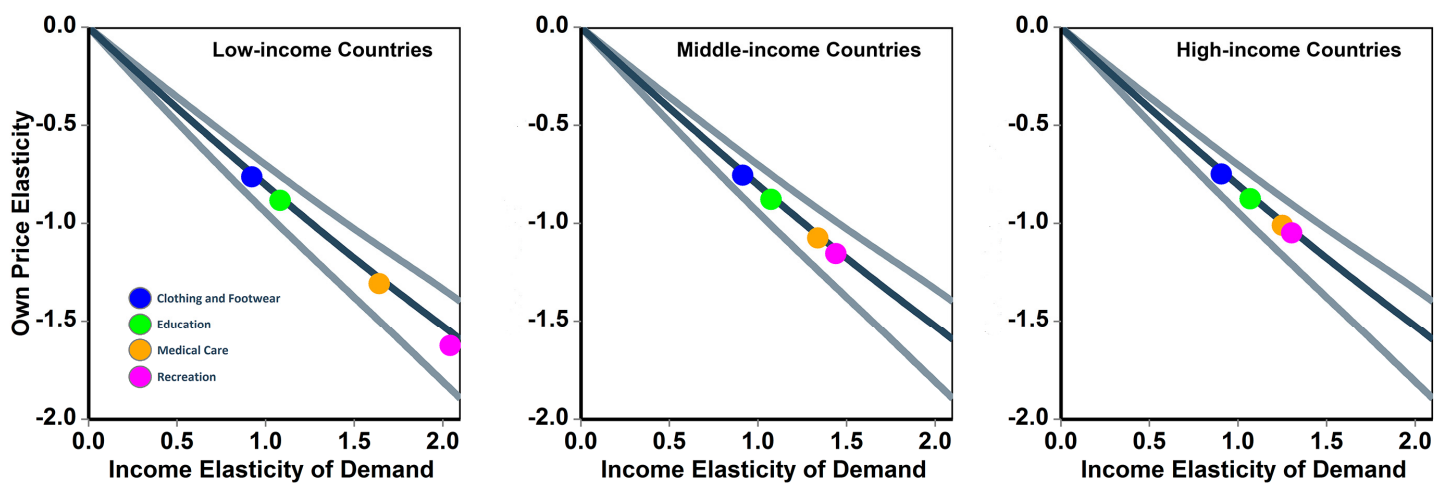

Fig 4. Comparison of simulation results with published estimates of income elasticity and uncompensated own price elasticity of demand for clothing and footwear, education, healthcare, and recreation. The estimates (colored circles) in the three panels refer to low-income, middle-income, and high-income countries and were obtained fitting the Florida model to country survey data (source: Seale JL, Regmi A, Bernstein J. International evidence on food consumption patterns. Economic Research Service; US Department of Agriculture; 2003). The darker line (in the middle) indicates the median of simulated values, while the lighter external lines define the 95% credible interval calculated using a Monte-Carlo simulation. The average budget share was drawn from a uniform distribution ranging from 0.0001 to 0.1, and the elasticity of the marginal utility of income was drawn from a normal distribution with mean equal to -1.26 and standard deviation equal to 0.1.

Table 1. Definitions

| Concept | Definition | Symbol |
|---|---|---|
| Utility function | a positive defined, functional relationship between purchased quantities ($q$) of market goods and the welfare (utility) of consumers | $u(q)$ |
| Marginal utility of income | the partial first derivative of the utility function with respect to income | $\mu$ |
| Elasticity of the marginal utility of income | the partial first derivative of the logarithm of the marginal utility of income, with respect to the logarithm of income | $\rho$ |
| Marginal share of a good | the partial first derivative of the share of consumer-income allocated to the purchase of a good (or bundle of goods) with respect to the consumer-income | $\theta$ |
| Consumer demand | the quantity of a market good (or of a bundle of goods) that the consumer purchases | $q$ |
| Income elasticity of demand | the partial first derivative of the logarithm of the Demand for a given good (or bundle of goods), with respect to the logarithm of income | $\varepsilon$ |
| Uncompensated own price elasticity of demand | the partial first derivative of the logarithm of the Demand for a given good (or bundle of goods) with respect to the logarithm of the own price of the good (or the average price of a bundle of goods) | $\eta$ |
| Uncompensated cross price elasticity of demand | the partial first derivative of the logarithm of the Demand for a good (or bundle of goods) named A with respect to the logarithm of the price of a good (or the average price of a bundle of goods) named B | $\psi$ |
| Preference independence | the condition that the utility associated with the consumption of goods belonging to a given bundle does not depend on the consumption of goods belonging to a different bundle | *No symbol* |